# Peripheral neuron survival and outgrowth on graphene

*Domenica Convertino[†,§], Stefano Luin[†], Laura Marchetti[\*,§], and Camilla Coletti[\*,§]*

[†]NEST, Scuola Normale Superiore, Piazza San Silvestro 12, 56127 Pisa, Italy

[§]Center for Nanotechnology Innovation @NEST, Istituto Italiano di Tecnologia, Piazza San Silvestro 12, 56127 Pisa, Italy

**E-mail:** camilla.coletti@iit.it, laura.marchetti@iit.it

## Abstract

Graphene displays properties which make it appealing for neuroregenerative medicine, yet its interaction with peripheral neurons has been scarcely investigated. Here, we culture on graphene two established models for peripheral neurons: PC12 cells and DRG primary neurons. We perform a nano-resolved analysis of polymeric coatings on graphene and combine optical microscopy and viability assays to assess the material cytocompatibility and influence on differentiation. We find that differentiated PC12 cells display a remarkably increased neurite length on graphene (up to 35%) with respect to controls. DRG primary neurons survive both on bare and coated graphene and present dense axonal networks.

## 1.Introduction

A specific feature of peripheral nerves is the ability to spontaneously regenerate after traumatic injuries. In the presence of important gaps where an end-to-end suture is not possible, a surgical approach is used, where nerve conduits (generally, autografts or allografts) are used as bridges between the nerve stumps and provide physical guidance for the axons (1). However, they present limitations in functional recovery and other disadvantages, *e.g.* size mismatch and increasing healing time for autografts, and rejection and disease transmission for allografts (2). A promising alternative is represented by tissue engineered nerve grafts, that have shown to improve regeneration, reduce scar formation and increase the concentration of neurotrophic factors (1,3). Among materials that can be used for the guide production, silicon stimulates excessive scar tissue formation thus lacking long-term stability, while some other natural polymers, such as collagen and chitosan, lack adequate mechanical and electrical properties (4–6). In recent years, new materials have been suggested as alternative candidates for tissue engineering applications. In particular graphene and other carbon-based nanomaterials have been proposed in life-science applications and nerve tissue regeneration (5,7,8).

Graphene is a monolayer of sp2-hybridized carbon atoms arranged in a two-dimensional honeycomb lattice that was first isolated in 2004 from graphite (9). The increasing research interest in graphene is due to its incredible properties: high electron mobility (also at room temperature), superior mechanical properties both in flexibility and strength, high thermal conductivity and high area/volume ratio (10,11). Furthermore, its biocompatibility and chemical stability make it ideally suited for biomedical applications (12).

Several studies have used graphene-based materials as biocompatible substrates for growth and differentiation of different cell types, including neural cells (13–16). To date, however, most studies have investigated graphene covalent functionalized forms such as graphene oxide (GO) and its chemical reduction known as reduced graphene oxide (RGO), or liquid phase exfoliated graphene. These graphene-like structures have altered electronic structure and physical properties due to the variable fraction of sp2 and sp3 hybridized carbon atoms. With respect to those graphene-based materials, pristine graphene offers enhanced electrical and tribological properties and most notably an excellent electrical conductivity thus prospecting advantages for nervous system regeneration applications. Indeed, it has been demonstrated that electrical stimulation enhances and directs neurite outgrowth (17,18).

To date, the interaction between pristine graphene and peripheral neural cells has been investigated only in two studies (19,20), which suggest a positive effect on neurite outgrowth and proliferation when using graphene coated with fetal bovine serum (FBS). However, in both studies bare glass is used as control, thus the effect on the results of FBS coating, which per se is not a traditional coating for neural cells (21), is not investigated. No detailed study has yet investigated the homogeneity and quality of the coatings typically adopted in neuronal culture. Predicting how polymeric surface coatings distribute onto graphene, due to its hydrophobicity and extreme flatness, is by no means trivial; furthermore, understanding how nerve cells can sense graphene under extracellular-matrix-like coatings is crucially important for possible *in vivo* applications.

Overall, this lack of studies on pristine graphene leaves other carbon-based materials such as carbon nanofibers (CNF), carbon nanotubes (CNT), GO and rGO to star in its play (5,8,15,22).

In this work we investigate the potential of graphene as a conductive peripheral neural interface. We select epitaxial graphene obtained via thermal decomposition on silicon carbide (SiC) (23) as the ideal substrate for such investigations. In fact, epitaxial graphene on SiC combines high crystalline quality, scalability, thickness homogeneity and an extreme cleanliness. Graphene is used as a substrate for two cellular models: (i) PC12 cells, a non-neuronal cell line that is able to differentiate upon Nerve Growth Factor (NGF) stimulation and constitutes a widely-used model for peripheral sympathetic neurons (24); (ii) dorsal root ganglion (DRG) sensory neurons, which are used as a model to study regenerative axon growth (25). The homogeneity and quality of a number of polymeric coatings typically adopted for neuronal culturing is investigated, and the most suitable ones are identified and adopted for the reported cultures. Optical microscopy is used to investigate neurite length, number and differentiation while viability assays are used to assess cytocompatibility. We compared results on monolayer graphene on SiC (G) with the ones on 4 possible control substrates: hydrogen etched SiC (SiC), gold coated glass coverslip (Au), glass coverslip (Glass) and polystyrene plate (well). The last two, being routinely used in cell culture procedures, were used as classic controls. SiC controls were implemented since graphene was grown directly on such substrates, which display a good biocompatibility (26) and present prospects for neural implants (27). Finally, gold substrates were used as conductive controls, as they can accelerate axonal elongation applying electrical stimulation (5).

**2. Materials and Methods**

*2.1. Substrates preparation and characterization*

Graphene on SiC was prepared by adopting a technique which allows to obtain quasi-free standing monolayer graphene (QFMLG) (28). Briefly, buffer layer graphene was obtained via thermal decomposition of on-axis 4H-SiC(0001) performed at 1250 °C in argon atmosphere. QFMLG was obtained by hydrogen intercalating the buffer layer samples at 900 °C in molecular hydrogen at atmospheric pressure (29). The controls adopted in the experiments were: (i) Hydrogen etched SiC(0001) dices – the same substrates were graphene was grown –were cleaned with HF to remove the oxide layer, and hydrogen etched at a temperature of 1250 °C as previously reported (30). (ii) Gold coated glass coverslips were obtained by thermally evaporating on the coverslips, previously cleaned with oxygen plasma, a 2 nm titanium adhesive layer and a 4 nm thin gold layer. (iii) Bare glass coverslips were treated overnight with 65% nitric acid (Sigma). The topography of the samples as well as the graphene number of layers and quality were assessed by both AFM and Raman spectroscopy (Figure S1).

*2.2. Surfaces functionalization*

Samples were coated with different polymeric solutions suggested for the targeted cell cultures and AFM analyses were performed to investigate the morphology of such coatings on graphene and the controls. The following solutions were tested: 0.1 % (w/v) Poly-L-lysine (PLL) solution in H2O (Sigma), 200 µg/ml Collagene Type I (Sigma) in deionized (DI) water, 30 µg/ml Poly-D-lysine (PDL) (Sigma) in PBS, 30 µg/ml PDL and 5 µg/ml laminin (Life Technologies) in PBS. The samples were incubated with the coating solution at 37 °C for 1 hour, 4 hours or 12 hours and rinsed three times in DI water before analyzing their topography via AFM. AFM was performed in tapping mode on samples with and without the polymeric coating, over several areas up to 10x10 µm wide. AFM micrographs were analyzed using the software Gwyddion 2.45.

*2.3. PC12 cell culture*

PC12 cells (ATCC® CRL-1721™) were maintained in a humidified atmosphere at 37 °C, 5% CO2 in RPMI 1640 medium supplemented with 10% horse serum, 5% fetal bovine serum, 1% penicillin/streptomycin and 1% L-glutamine (Gibco). Cells were plated at ~40-60% confluency onto the substrates previously coated with 0.1 % (w/v) Poly-L-lysine solution (PLL) in H2O (Sigma). Differentiation was achieved using two different procedures: 1) direct addition of 50 ng/ml NGF (Alomone Labs) in complete cell medium after seeding; 2) a 5-6 days priming with 15 ng/ml NGF in complete medium, followed by seeding on the substrates with 50 ng/ml NGF in RPMI medium supplemented with 1% horse serum, 0.5% fetal bovine serum, 1% penicillin/streptomycin and 1% L-glutamine. In both cases, 2/3 of the medium was renewed every 2-3 days. With the second procedure an improved differentiation was observed. The cells were observed at different time points using an inverted microscope equipped with a 20x/40x magnification objective (Leica DMI4000B microscope). Typically, 10 fields per sample were acquired to perform morphometric analysis of PC12 differentiation. Three parameters were measured as previously reported (31): (i) the percentage of differentiated cells (Diff), determined counting the number of cells with at least one neurite with a length equal to or longer than the cell body diameter; (ii) the average number of neurites per cell in the field (av. neurites/cell); (iii) the mean neurite length measuring the longest neurite of each differentiated cell in the field.

Cell viability was assessed with the Cell counting Kit-8 assay (CCK-8, Sigma), based on quantification of WST reduction due to the metabolic activity of viable cells. Samples were prepared according to the manufacturer's instructions and measured at the GloMax® Discover multiplate reader (Promega). The results are reported as % over the polystyrene well, considered as control. All the experiments were repeated at least twice independently.

*2.4. DRG Cell Culture*

Rat Embryonic Dorsal Root Ganglion Neurons (R-EDRG-515 AMP, Lonza) cells were maintained in a humidified atmosphere at 37 °C, 5% $CO_2$ in Primary Neuron Basal Medium (PNBM, Lonza) supplemented with L-glutamine, antibiotics and NSF-1 (at a final concentration of 2%) as recommended by the manufacturer. Neurons were plated on the substrates previously coated with a PBS solution of 30 μg/ml Poly-D-lysine (Sigma) (PDL) and 5 μg/ml laminin (Life Technologies). The medium was always supplemented with 100 ng/ml of NGF (Alomone Labs). Since 24h after seeding, 25 μM AraC (Sigma) was added for inhibition of glia proliferation. Half of the medium was replaced every 3-4 days. Neurons were observed at different time points using an inverted microscope (Leica DMI4000B microscope).

**3. Results and discussion**

*3.1. Polymeric coating of epitaxial graphene and control substrates*

NGF-induced neurite outgrowth of PC12 cells is favored by their adhesion on a substrate. This is typically achieved by coating the dish surfaces with polymers such as poly-L-lysine or biologically derived collagen (24). We applied a water solution of both these coatings to all substrates adopted for our cultures and analyzed by AFM the quality and homogeneity of the coatings after different incubation times, *i.e.* 1 hour, 4 hours and 12 hours. Panels (a) and (b) of **Figure 1** show AFM phase and topography micrographs for the two different coatings and different incubation times on a graphene substrate. Clearly, the Poly-L-Lysine (PLL) coating presents better homogeneity with respect to Collagen Type I coating for which network-like aggregates can be detected. On the other hand, PLL tends to form a homogeneous carpet of spots of 1-2 nm (no aggregates) independent from the incubation time. We also analyzed the same coatings on SiC, gold and glass surfaces. On SiC, PLL and Collagen presented analogous topographies (Figure S2(a) and (b)). Due to the high surface roughness of gold and glass substrates, no conclusions about the quality of the coating could be drawn (Figure S3), although its presence was confirmed by the variation in the hydrophilicity observed with contact angle measurements (Figure S4). Hence, for the PC12 cells cultured in this work, a PLL coating with an incubation time of 4 hours was adopted.

The same characterization was performed for the polymeric coatings typically suggested for DRG neurons, i.e., PBS solution of Poly-D-Lysine (PDL) alone and PDL with laminin. Panel (c) in Figure 1 shows the AFM topography and phase images taken for PDL/laminin coated graphene substrates for the three different incubation times (i.e., 1 hour, 4 hours and 12 hours). Also in this case, after the coating, an increased roughness was observed for all time points and in particular the formation of a network-like structure was consistently observed. PDL alone coating gave rise to a similar net (Figure S5(b)). In order to exclude the effect of PBS, we dissolved the same polymeric amount in DI water and after 4h incubation we observed similar structures (Figure S5(a)). On SiC no network formation was observed with or without laminin (Figure S2(c) and (d)). The stability of the coating was confirmed for all the probed incubation times. In this case, PDL with laminin coating (with an incubation time of 4 hours) was selected to carry on the following DRG culture experiments in order to mimic the extracellular matrix.

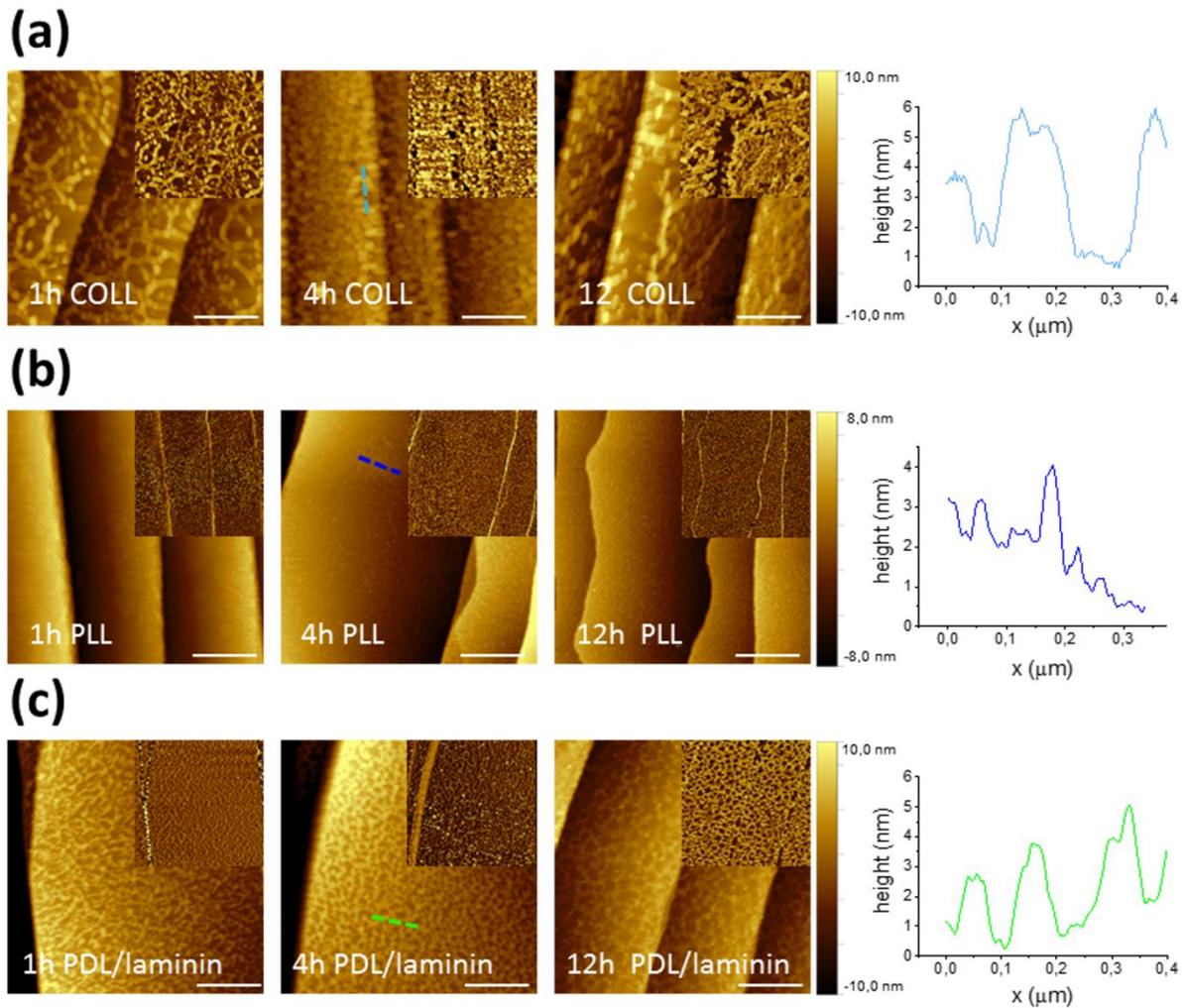

Figure 1. AFM topography images with characteristic line profiles of graphene after three different times of incubation (1 hour, 4 hours and 12 hours) with Collagen Type I coating (200 μg/ml in DI water) (a), 0.1 % (w/v) Poly-L-lysine solution in H2O (b) and Poly-D-Lysine and Laminin coating (30 μg/ml PDL and 5 μg/ml laminin in PBS) (c) (scale bar: 500 nm). The insets show phase images of the same areas. AFM line profile after 4 hours-incubation are shown for each coating.

*3.2. Neurite outgrowth of PC12 cell on graphene*

We first investigated the effect of graphene on PC12 cells. **Figure 2**(a) reports typical optical micrographs obtained for PC12 cells cultured at day 5 (in the presence and absence of NGF) and at day 7 (with NGF) on the different substrates. The analyses conducted at day 5 evidence that almost no differentiation took place in the absence of NGF, while a significant neurite outgrowth occurred on all substrates upon NGF treatment.

Selected morphometric parameters describing the differentiation process were quantified at day 5 and are reported in panels (b), (c) and (d) of Figure 2: the percentage of differentiated cells in the fields (Diff), the average number of neurites per cell (av. neurites/cell) and the length of the longest neurite per differentiated cell (length). This analysis showed that 50% of the cells on graphene differentiate with a mean neurite length of 48.7 µm (panels (b) and (c)). Remarkably, the average length was significantly longer on graphene than on glass (***) and well (**) by 35% and 22% respectively. No significant difference was instead observed in the percentage of differentiation and in the average number of neurites per cell. These results indicate that PC12 cells grow longer neurites on graphene, with a neuronal differentiation that is comparable to that obtained for the standard control wells. Differently from reference (20), we did not observe increased PC12 proliferation on graphene, which could be due to the effect of the FBS coating used in that study. Furthermore, we found that at day 7 living PC12 cells forming neurite networks were present on all the substrates. To better assess graphene cytocompatibility, the viability of undifferentiated PC12 cells was assessed after 3, 5 and 7 days of culture and no statistically significant differences were observed between graphene and the other substrates (Figure 2(e)).

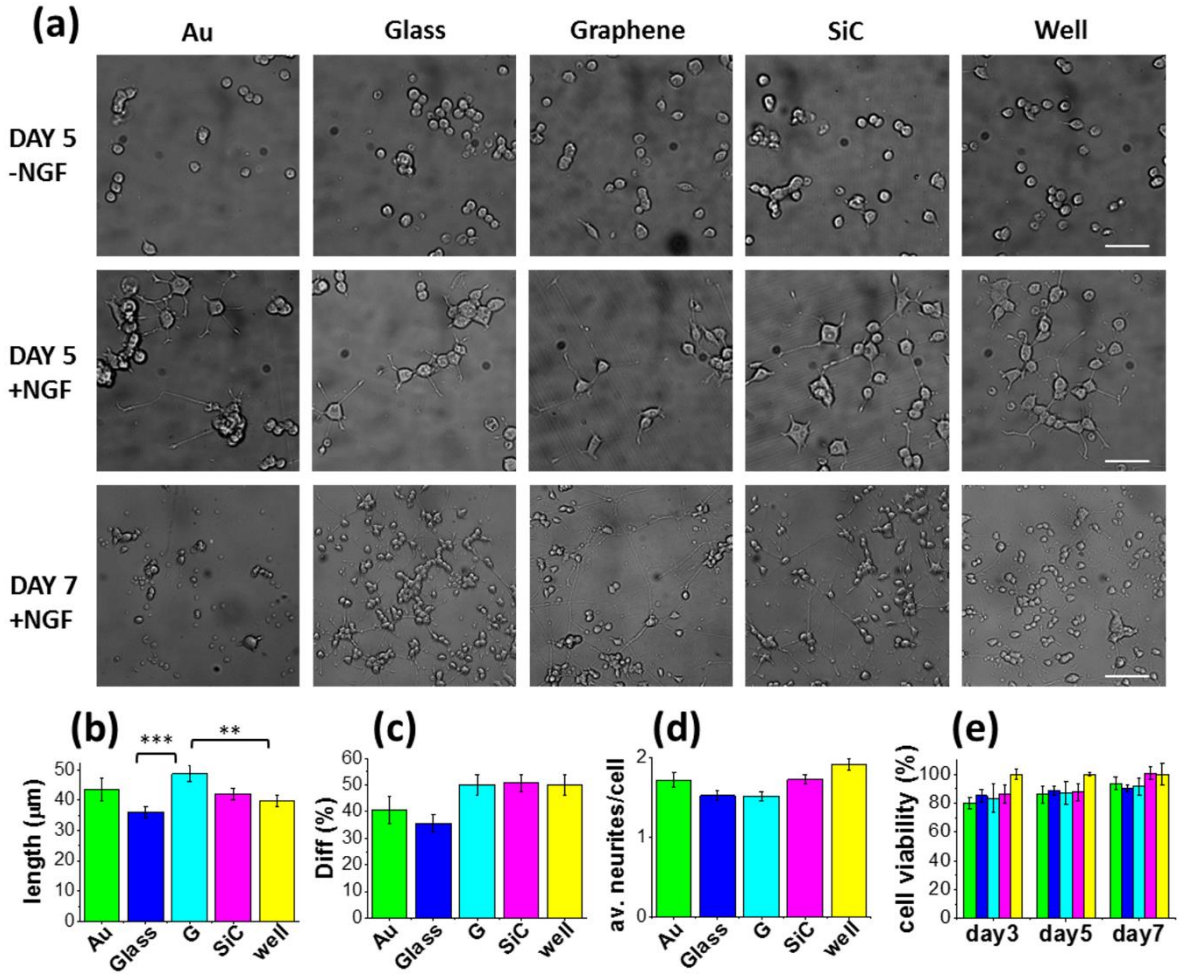

Figure 2. (a) PC12 grown on gold (Au), glass coverslip (Glass), graphene (G), SiC and polystyrene (well) coated with PLL in the absence of NGF (first row, scale bar: 50 μm), PC12 cells differentiation at day5 (second row, scale bar: 50 μm) and day 7 (third row, scale bar: 100 μm). Histograms show the quantification of (b) neurite length, (c) percentage of differentiation and (d) average number of neurites per cell after 5 days of NGF treatment of two independent experiments per substrate. For each substrate we analyzed at least 200 cells (nc) from selected fields (nf) (Au: nf=17, nc=203; Glass: nf=20, nc=798; G: nf=20, nc=442; SiC: nf=20, nc=607; well: nf=20, nc=527). (e) Cell viability after 3, 5 and 7 days tested by WST-8. Results are expressed as percentage of cell proliferation relative to the proliferation on the polystyrene control sample. Bars colored as in the other graphs. Data reported as mean ± SE. Nonparametric Kruskal–Wallis test was used for statistical significance, with ** $p<0.01$, *** $p<0.001$.

*3.3. DRG primary neurons on graphene*

Next, we investigated the effect of graphene on primary neurons using dorsal root ganglion (DRG) cells while using the same controls adopted in the previous culture. As motivated in 3.1, all the samples were coated with PDL/laminin. **Figure 3**(a) shows typical optical microscopy images obtained at 1, 4, 9 and 15 days of culture. Starting from day 4, we observed numerous processes and an increase in the cell body area (Figure S6) and in the neurite length (Figure 3(a)). Neurons were observed on all the substrates up to 17 days of culture. We observed that both at day 1 and day 2 the average axon length was higher on graphene than on the other substrates (Figure 3(b)). This observation confirms the trend reported for PC12, although in this case no statistical significance was retrieved. Axonal length was not quantified for longer culturing times due to the highly dense network forming after day 2 (see day 9 and 15 in Figure 3(a)).

Given that neuronal growth was previously reported also for non-coated graphene (32,33), we tested also the bare substrates to observe their effect on the neurons. Differently from non-coated glass, where they did not survive, DRG neurons could be nicely cultured on non-coated graphene and gold. On these uncoated substrates, DRG formed cell bodies aggregates and neurite bundles (Figure 3(c)), as previously observed for retinal ganglion cells cultured on graphene (32). Remarkably, DRG neurons survived on uncoated graphene and gold up to 17 days.

Concerning material stability issues, it should be noted that graphene showed a good stability and remained intact during the entire culturing period, as revealed by Raman measurements after cell removal (Figure S7).

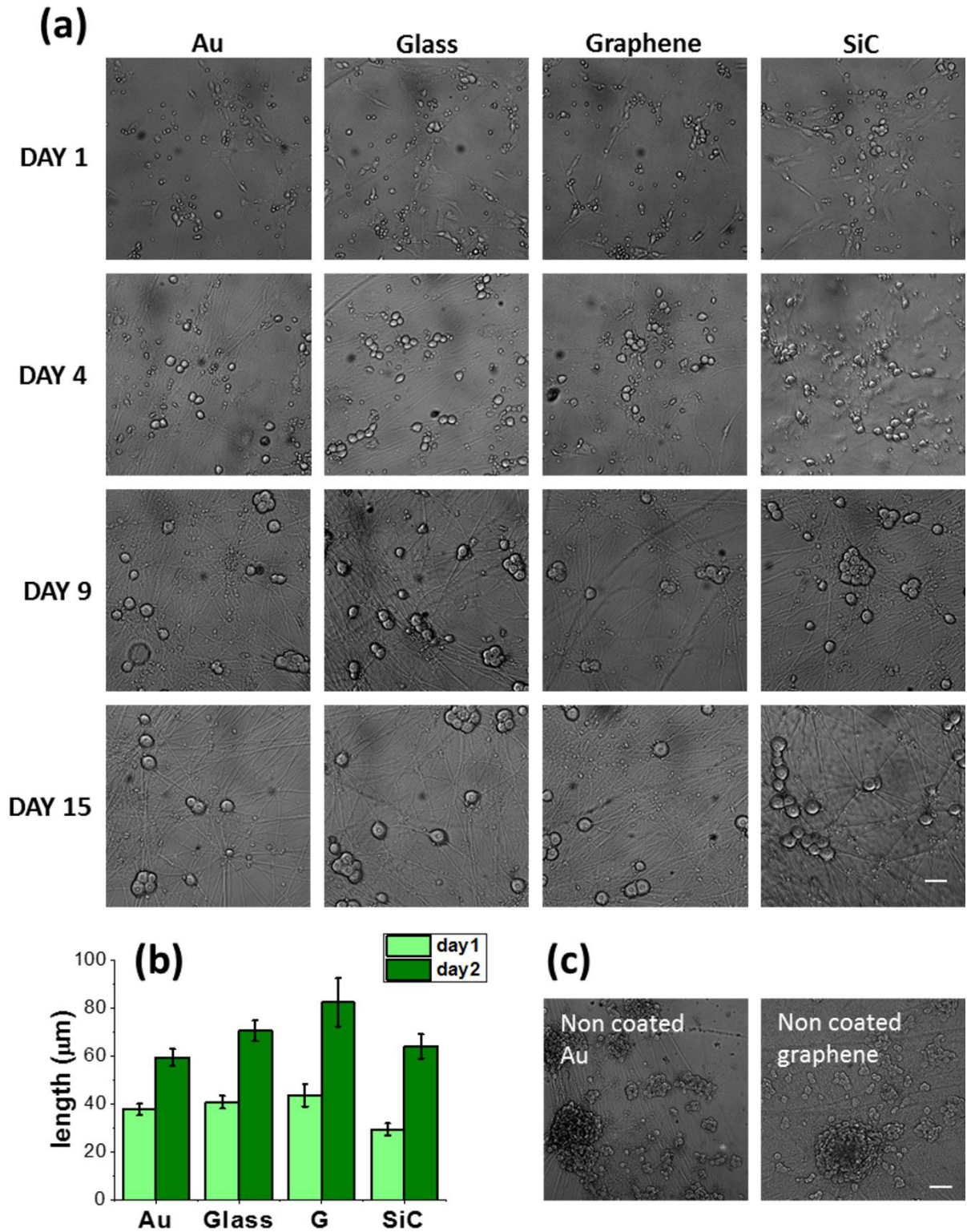

Figure 3. (a) DRG neurons cultured on gold (Au), glass coverslip, graphene (G) and SiC coated with Poly-D-lysine and laminin at different days of culture. Scale bar: 50 µm. (b) Axon length quantification at 24 and 48 hours after cell seeding. We analyzed nf fields for a total of nc cells for each substrate (day1: Au, nf =13, nc=67, Glass: nf =14, nc=75; G: nf =13, nc=29; SiC: nf =12, nc=35; day2: Au, nf=16, nc=89, Glass: nf =13,

nc=100, G: nf=12, nc=34, SiC: nf =11, nc=37) and data are reported as mean ± SE. (c) DRG neurons on bare gold and graphene at day 10. Scale bar: 100 μm.

## 4. Conclusion

This work provides novel data about the use of graphene as a substrate for peripheral neuron cultures. We use the PC12 cell line as a consolidated model for peripheral sympathetic neurons and show that such cells grow well on graphene with an increased neurite length (up to 35%) at 5 days of differentiation when compared to controls. Remarkably, graphene performs better than gold, which is an appealing conductive candidate for biomedical applications. Culture of DRG neurons also shows a positive outcome on graphene: neurons survive both on bare and coated graphene until day 17, with a dense axon network that is comparable to the control substrates. In order to investigate graphene influence on axonal outgrowth, further studies are necessary, *e.g.* using compartmentalized chambers (34). The obtained results confirm the potential of graphene as an active substrate in nerve guidance conduit devices: it would allow the transmission of electrical signals between neurons and make external electrical stimulation feasible to enhance axon regeneration.

## Acknowledgment

The authors would like to thank Fabio Beltram and Giovanni Signore from NEST-Scuola Normale Superiore for fruitful discussion. This project has received funding from the European Union's Horizon 2020 research and innovation programme under grant agreement No. 696656-GrapheneCore1.

# Supporting Information

# Peripheral neuron survival and outgrowth on graphene

*Domenica Convertino[†,§], Stefano Luin[†], Laura Marchetti[*,§], and Camilla Coletti[*,§]*

[†]NEST, Scuola Normale Superiore, Piazza San Silvestro 12, 56127 Pisa, Italy

[§]Center for Nanotechnology Innovation @NEST, Istituto Italiano di Tecnologia, Piazza San Silvestro 12, 56127 Pisa, Italy

E-mail: camilla.coletti@iit.it, laura.marchetti@iit.it

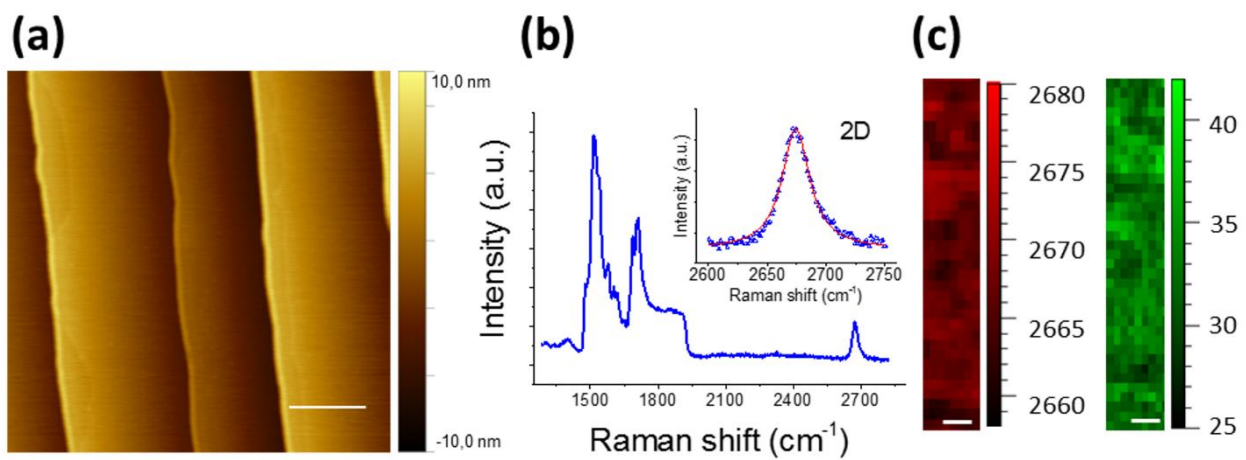

**Figure S1.** (a) Characteristic AFM topography of an intercalated graphene sample, showing atomically flat terraces separated by steps (scale bar: 400 nm). (b) Raman spectrum of an intercalated graphene sample, obtained using a 532 nm laser and a 50x objective lens. The insert shows the single Lorentzian fitting of the 2D peak, with a narrow FWHM of 28 cm-1. (c) 2D peak position (left) and FWHM (right) distribution in a large area (scale bar: 2 µm). The position and shape of the 2D (~2700 cm-1) peak, originated from a double resonance electron-phonon scattering process, give an indication of the doping and the number of graphene layers. In particular, the single Lorentzian fitting of the peak is characteristic of monolayer graphene, while for bilayer and trilayer graphene the 2D peak becomes broader and the fitting requires multiple Lorentzians. The energy of the peak, blue-shifted with respect to the case of pure undoped graphene, indicates a p-type doping, characteristic of a quasi-free standing monolayer graphene (QFMLG).

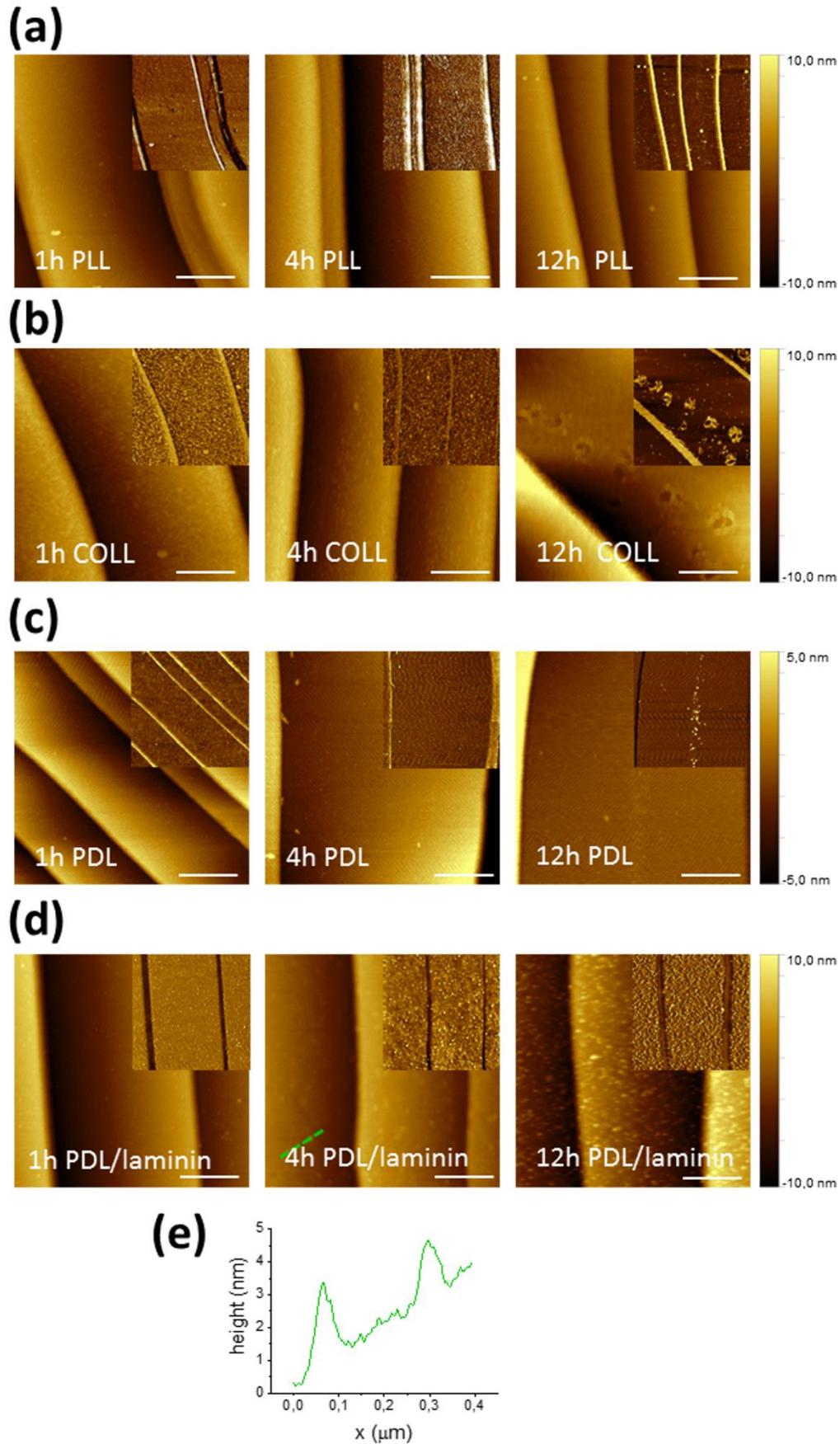

**Figure S2.** AFM topography images of SiC samples after three different times of incubation (1 hour, 4 hours or 12 hours) with a coating solution of: (a) PLL, (b) collagen, (c) PDL, (d)

PDL/laminin (scale bar: 500 nm). The insets show phase images of the same areas, which are not sensitive to slow changes in height and improve identification of nanometric structures. (e) All the samples are coated with a homogeneous carpet of spots of few nanometers, as showed in the AFM line profile of a SiC sample after 4 hours incubation with PDL/laminin.

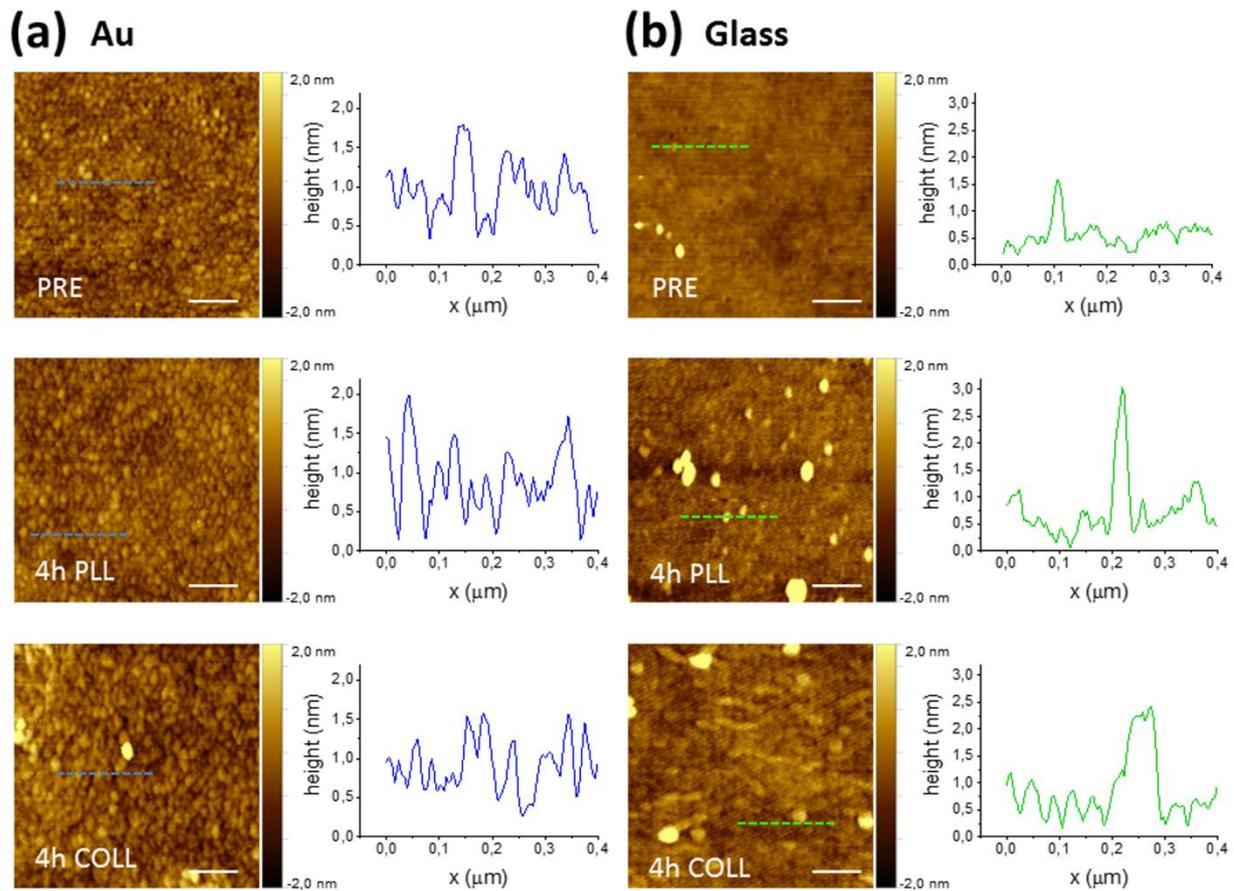

**Figure S3.** AFM topography and roughness profiles of gold (a, Au) and nitric acid treated glass (b, Glass) before protein coating and after 4h incubation with Poly-L-lysine (4h PLL) and Collagen Type I (4h COLL) (scale bar: 200 nm). Both the surfaces revealed an initial roughness comparable to the one after the coating, preventing the recognition of nanometric details.

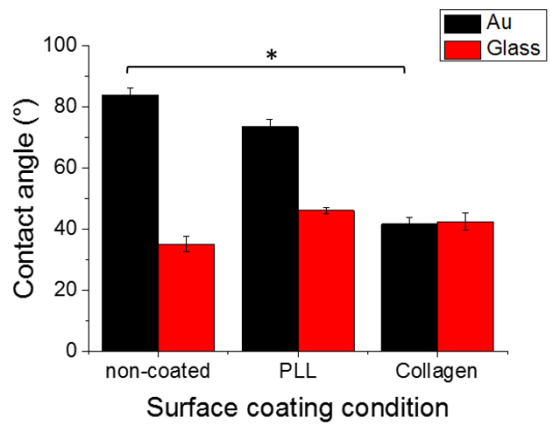

**Figure S4.** Contact angle measurements of gold (Au) and nitric acid treated glass (Glass) before protein coating and after 4h incubation with Poly-L-lysine (PLL) and Collagen Type I (Collagen). All measurements were made using DI water as a probe liquid. Values are the mean ± standard deviation for 3 samples. Non-coated gold was more hydrophobic than non-coated glass. The coatings had opposite effects on the substrates, increasing hydrophilicity for gold and increasing hydrophobicity for glass.

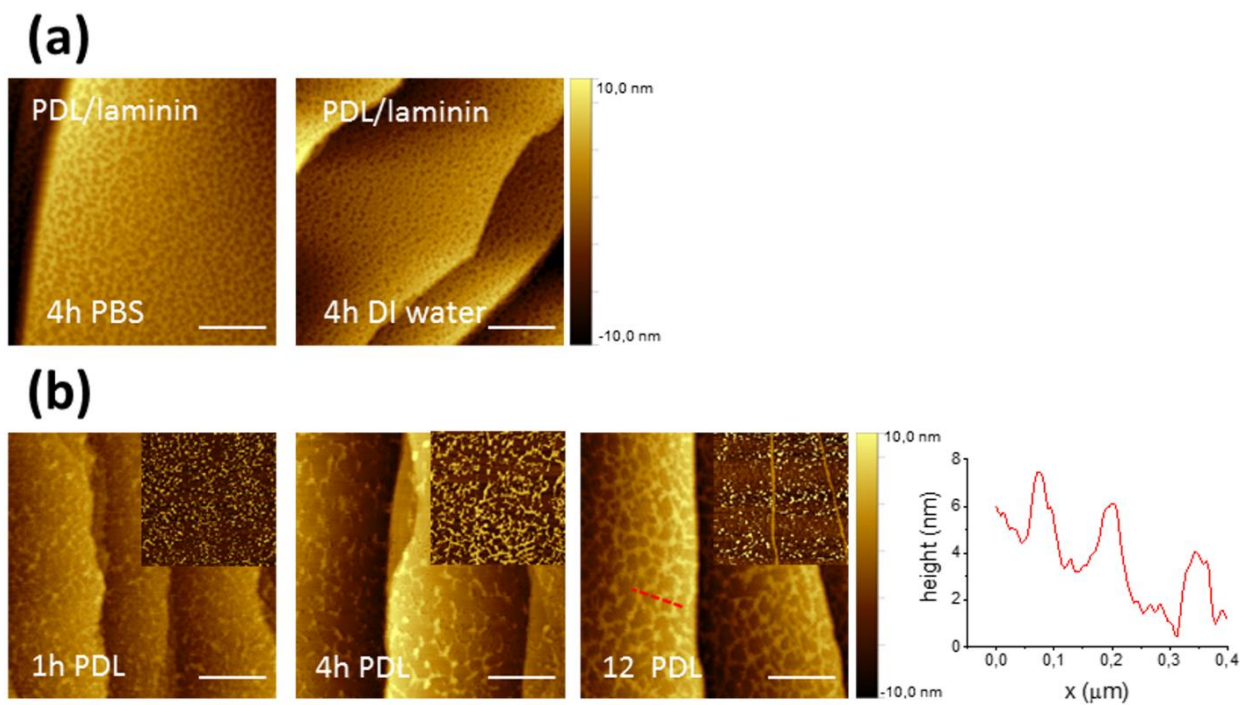

**Figure S5.** (a) AFM topography of graphene samples coated with PDL/laminin dispersed in DI water and PBS after 4h incubation show similar net structures. This implies that the net morphology is independent from the salts in the PBS solution. (b) AFM topography images with a characteristic line profiles of graphene after three different times of incubation (1 hour, 4 hours and 12 hours) with Poly-D-Lysine (PDL) coating (scale bar: 500 nm).

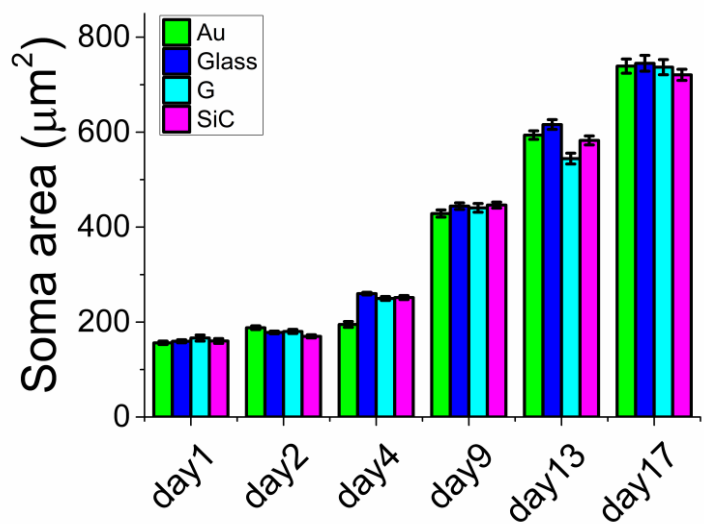

**Figure S6.** Increasing of the cell body area with time in dorsal root ganglion (DRG) cells. For cell soma analysis more than 100 cells per sample were analysed. Cell bodies were approximated to an oval shape and relative areas were evaluated using ImageJ.

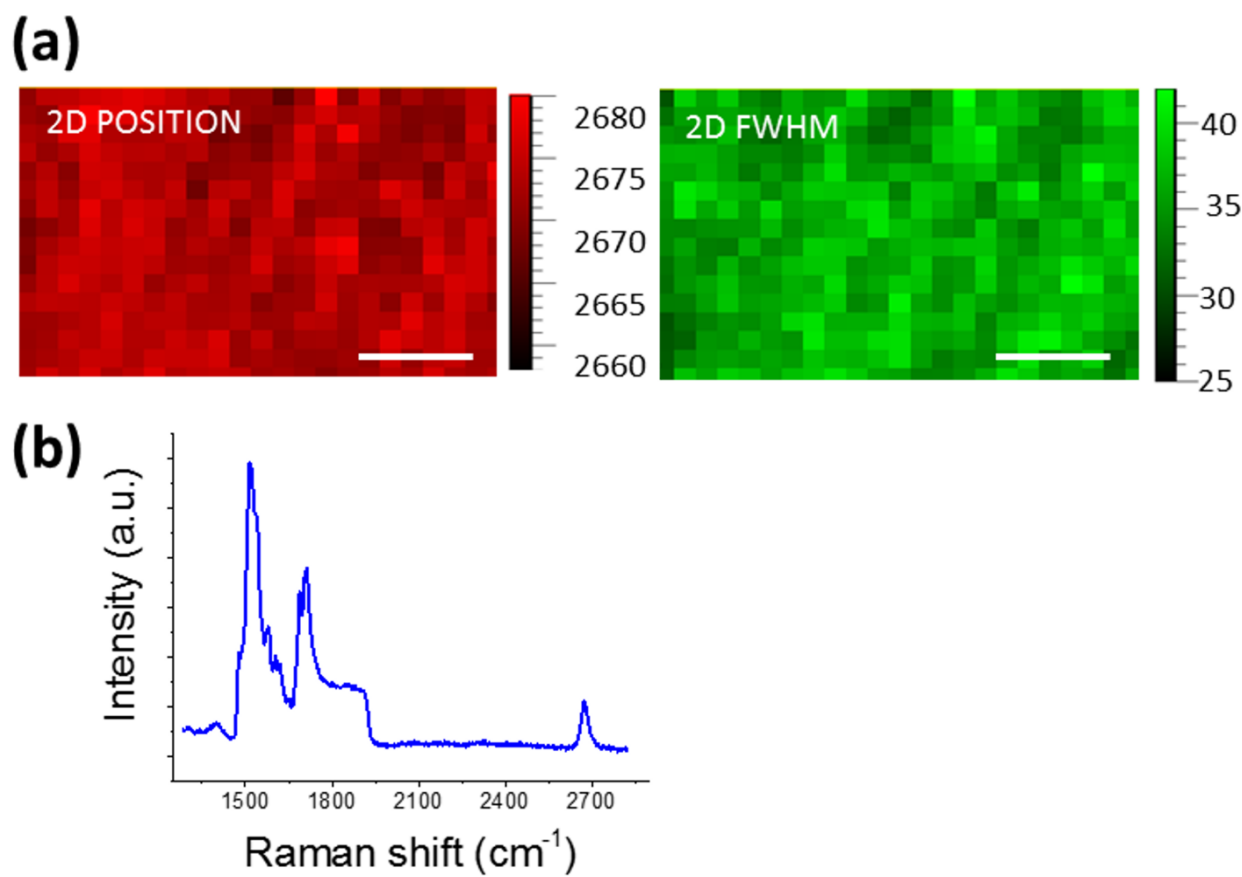

**Figure S7.** Raman characterization with 532 nm laser of a graphene sample after cell culture validates the full coverage of graphene. (a) 2D peak position and FWHM distribution in a large area (scale bar: 5 μm). (b) Characteristic Raman spectrum. The maps reveal that the 2D peak and FWHM are very homogeneous across the whole area and the values resemble those measured before the cell culture, with a narrow 2D peak of ~30 cm-1 centered at ~2670 cm-1.

*Statistical analysis*

All data are expressed as the average value (mean) ± standard error of the mean (SE) unless stated otherwise. Data were analyzed by using Origin Software and nonparametric Kruskal–Wallis test was used for statistical significance with *p<0.05, ** p<0.01 and *** p<0.001.